\documentclass[acmtocl,acmnow]{acmtrans2m}

\newtheorem{theorem}{Theorem}[section]

\newtheorem{lemma}[theorem]{Lemma}
\newtheorem{fact}[theorem]{Fact}
\newtheorem{claim}[theorem]{Claim}
\newdef{definition}[theorem]{Definition}
\newdef{remark}[theorem]{Remark}
\newdef{problem}[theorem]{Problem}

\usepackage{latexsym}
\usepackage{amssymb,amsmath}
\usepackage{amsfonts}

\newcommand{\s}
{\mbox{\boldmath $s$}}

\newcommand{\bp}
{\mbox{\boldmath $P$}}

\newcommand{\ml}{\sf }

\newcommand{\Ax}{\textit{Nom}}

\title{Undecidability of the unification and admissibility problems for modal
and description logics}
\author{Frank Wolter\\University of Liverpool
\and Michael Zakharyaschev\\
Birkbeck College London}

\begin{abstract}
We show that the unification problem `is there a substitution instance of a
given formula that is provable in a given logic?' is undecidable for basic modal
logics {\ml K} and {\ml K4} extended with the universal modality. It follows
that the admissibility problem for inference rules is undecidable for these
logics as well. These are the first examples of standard decidable modal logics
for which the unification and admissibility problems are undecidable. We also
prove undecidability of the unification and admissibility problems for {\ml K}
and {\ml K4} with at least two modal operators and nominals (instead of the
universal modality), thereby showing that these problems are undecidable for
basic hybrid logics. Recently, unification has been introduced as an important
reasoning service for description logics. The undecidability proof for {\ml K}
with nominals can be used to show the undecidability of unification for Boolean
description logics with nominals (such as $\mathcal{ALCO}$ and
$\mathcal{SHIQO}$). The undecidability proof for {\ml K} with the universal
modality can be used to show that the unification problem relative to role boxes
is undecidable for Boolean description logics with transitive roles, inverse
roles and role hierarchies (such as $\mathcal{SHI}$ and $\mathcal{SHIQ}$).
\end{abstract}

\category{F.4.1}{Mathematical logic and formal languages}{modal logic.}

\terms{theory.}

\keywords{unification, admissible rule, description logic, hybrid logic,
decidability.}

\begin{document}

\begin{bottomstuff}
Author's address: F.~Wolter, Department of Computer Science, University of
Liverpool, Liverpool L69 7ZF, U.K., {\tt frank@csc.liv.ac.uk}. M.~Zakharyaschev,
School of Computer Science and Information Systems, Birkbeck College, London
WC1E 7HX, U.K., {\tt michael@dcs.bbk.ac.uk}.
\newline
\end{bottomstuff}

\maketitle


\section{Introduction}

The \emph{unification} (or \emph{substitution}) \emph{problem} for a
propositional logic $L$ can be formulated as follows: given a formula $\varphi$
in the language of $L$, decide whether it is \emph{unifiable} in $L$ in the
sense that there exists a uniform substitution $\s$ for the variables of
$\varphi$ such that $\s(\varphi)$ is provable in $L$. For normal modal logics,
this problem is equivalent to the standard unification problem modulo equational
theories \cite{Baader&Siekmann}: in this case the equational theory consists of
any complete set of equations axiomatising the variety of Boolean algebras with
operators and additional equations corresponding the axioms of $L$.

A close algorithmic problem for $L$ is the \emph{admissibility problem} for
inference rules: given an inference rule $\varphi_1,\dots,\varphi_n /\varphi$,
decide whether it is {\em admissible} in $L$, that is, for every substitution
$\s$, we have $L \vdash \s(\varphi)$ whenever $L\vdash\s(\varphi_1)$, \ldots,
$L\vdash \s(\varphi_n)$. It should be clear that if the admissibility problem
for $L$ is decidable, then the unification problem for $L$ is decidable as well.
Indeed, the rule $\varphi/\bot$ is not admissible in $L$ iff there is a
substitution $\s$ for which $L\vdash \s(\varphi)$.

It follows from the results of V.~Rybakov (see \cite{Rybakovbook97} and
references therein; see also
\cite{Ghilardi2000,Ghilardi2004,Ghilardi&Sacchetti2004,Iemhoff01,Iemhoff03})
that the unification and admissibility problems are decidable for propositional
intuitionistic logic and such standard modal logics as {\ml K4}, {\ml GL}, {\ml
S4}, {\ml S4.3}. However, nearly nothing has been known about the decidability
status of the unification and admissibility problems for other important modal
logics such as the (`non-transitive') basic logic {\ml K}, various multi-modal,
hybrid and description logics. In fact, only one---rather artificial---example
of a decidable \emph{uni}modal logic for which the admissibility problem is
undecidable has been found \cite{Chagrov92b} (see also \cite{Chagrov&Z97}).

\emph{The first main result of this paper shows that for the standard modal
logics {\ml K} and {\ml K4} {\rm (}and, in fact, all logics between them{\rm )}
extended with the universal modality the unification problem and, therefore, the
admissibility problem are undecidable.}

The universal modality, first investigated in \cite{Goranko&Passy92}, is
regarded nowadays as a standard constructor in modal logic; see, e.g.,
\cite{MLHandbook}. Basically, the universal box is an {\ml S5}-box whose
accessibility relation contains the accessibility relations for all the other
modal operators of the logic. The undecidability result formulated above also
applies to those logics where the universal modality is definable, notably to
propositional dynamic logic with the converse; see, e.g., \cite{Hareletal00}.
The unification and admissibility problems for {\ml K}
itself still remain open. Observe that {\ml K4} is an example of a logic for
which the unification and admissibility problems are decidable, but the addition
of the (usually `harmless') universal modality makes them undecidable (although
{\ml K4} with the universal modality itself is still decidable, in fact, {\sc
PSpace}-complete). Note also that for `reflexive' modal logics with the
universal modality such as {\ml S4} the unification problem is trivially
decidable.

\emph{The second result of this paper shows that the unification and
admissibility problems are undecidable for multimodal {\ml K} and {\ml K4} {\rm
(}with at least two modal operators{\rm )} extended with nominals.}

Nominals, that is, additional variables that denote singleton sets, are one of
the basic ingredients of hybrid logics; see, e.g., \cite{Arecesten} and
references therein. As follows from our second result, for most hybrid logics
the unification and admissibility problems are undecidable.

A particularly interesting consequence of this result is in description logic.
Motivated by applications in the design and maintenance of knowledge bases,
Baader and Narendran \citeyear{Baader&Narendran} and Baader and Kuesters
\citeyear{BaaderKuesters-LPAR} identify the unification problem for concept
descriptions as an important reasoning service. In its simplest formulation,
this problem is equivalent to the unification problem for modal logics. Baader
and Narendran \citeyear{Baader&Narendran} and Baader and Kuesters
\citeyear{BaaderKuesters-LPAR} develop decision procedures for certain
sub-Boolean description logics, leaving the study of unification for Boolean
description logics as an open research problem. It follows from our results that
unification is undecidable for Boolean description logics with nominals such as
$\mathcal{ALCO}$, $\mathcal{ALCQO}$, $\mathcal{ALCQIO}$, and $\mathcal{SHIQO}$.
Moreover, if a Boolean description logic has transitive roles, inverse roles and
role hierarchies, then a role box can be used to define a universal role. In
this case our results can be used to show the undecidability of unification
relative to role boxes. This applies, for example, to the logics $\mathcal{SHI}$
and $\mathcal{SHIQ}$. These undecidability results cover almost all Boolean
description logics used in applications, in particular the description logic
underlying {\ml OWL-DL}. However, the unification problem for some basic Boolean
description logics such as $\mathcal{ALC}$ and $\mathcal{ALCQI}$ remains open.

The plan of this paper is as follows. We start by introducing the syntax and
semantics of normal modal logics with the universal modality, in particular
${\ml K4}_{u}$ and ${\ml K}_{u}$. Then we prove, using an encoding of Minsky
machines, the undecidability of the unification and admissibility problems for
all logics between ${\ml K4}_{u}$ and ${\ml K}_{u}$. We also briefly discuss the
formulation of this result in terms of equational theories. Then we introduce
modal logics with nominals and show how to modify the proof in order to
establish the undecidability of unification and admissibility for ${\ml K}$
and ${\ml K4}$ with at least two modal operators and nominals. We close with a brief discussion of consequences
for description logics with nominals.


\section{Unification in modal logics with the universal
modality}\label{universal}

Let ${\cal L}$ be the propositional language with an infinite set
$p_{0},p_{1},\ldots$ of propositional variables, the Boolean connectives
$\wedge$ and $\neg$ (and their derivatives such as $\vee$, $\rightarrow$, and
$\bot$), and two unary modal operators $\Box$ and $\forall$ (with their duals
$\Diamond$ and $\exists$). A \emph{normal modal logic} $L$ with the
\emph{universal modality} $\forall$ is any set of ${\cal L}$-formulas that
contains all propositional tautologies, the axioms
\begin{align*}
& \Box (p \rightarrow q) \rightarrow (\Box p \rightarrow \Box q), \qquad \forall
(p
\rightarrow q) \rightarrow (\forall p \rightarrow \forall q),\\
& \forall p \rightarrow  p, \qquad \forall p \rightarrow \forall\forall p,
\qquad p \rightarrow \forall \exists p, \qquad \forall p \rightarrow \Box p,
\end{align*}
and is closed under \emph{modus ponens}, the necessitation rules $\varphi/\Box
\varphi$ and $\varphi/\forall \varphi$, and uniform substitution. ${\ml K}_{u}$
is the smallest normal modal logic with the universal modality. ${\ml K4}_{u}$
is the smallest normal modal logic with the universal modality that contains the
extra axiom $\Box p \rightarrow \Box\Box p$.

${\ml K}_{u}$ and ${\ml K4}_{u}$ as well as many other normal modal logics with
the universal modality are determined by relational structures. A \emph{frame}
for ${\cal L}$ is a directed graph ${\mathfrak F}=(W,R)$, that is, $R\subseteq
W\times W$. A \emph{model} for ${\cal L}$ is a pair ${\mathfrak M}= ({\mathfrak
F},{\mathfrak V})$ where ${\mathfrak F}$ is a frame and ${\mathfrak V}$ a
\emph{valuation} mapping the set of propositional variables to $2^{W}$. The
\emph{truth-relation} $(\mathfrak M,x)\models\varphi$ between points $x\in W$ of
$\mathfrak M$ and $\mathcal{L}$-formulas $\varphi$ is defined inductively as
follows:
\begin{itemize}
\item[] $(\mathfrak M,x) \models p_{i}$ iff $x\in {\mathfrak V}(p_{i})$,

\item[] $(\mathfrak M,x) \models \neg \psi$ iff $(\mathfrak M,x) \not\models
\psi$,

\item[] $(\mathfrak M,x) \models \psi \land \chi$ iff $(\mathfrak M,x) \models \psi$
and $(\mathfrak M,x) \models \chi$,

\item[] $(\mathfrak M,x) \models \Box\psi$ iff $(\mathfrak M,y) \models \psi$ for
all $y\in W$ with $xRy$,

\item[] $(\mathfrak M,x) \models \forall \varphi$ iff $(\mathfrak M,y) \models
\varphi$ for all $y\in W$.
\end{itemize}
Instead of $(\mathfrak M,x)\models\varphi$ we write $x\models\varphi$ if
$\mathfrak M$ is clear from the context.

A formula $\varphi$ is \emph{valid} in a frame ${\mathfrak F}$, ${\mathfrak
F}\models \varphi$ in symbols, if $\varphi$ is true at every point of every
model based on ${\mathfrak F}$. The following facts are well known (see, for
example, \cite{Arecesetal00}):

\begin{fact}
${\ml K}_{u}$ is the set of formulas that are valid in all frames. ${\ml
K4}_{u}$ is the set of formulas that are valid in all transitive frames. The
satisfiability problem is {\sc ExpTime}-complete for ${\ml K}_{u}$, and {\sc
PSpace}-complete for ${\ml K4}_{u}$.
\end{fact}

We now formulate the unification problem for normal modal logics with the
universal modality.

\begin{definition}
The \emph{unification problem} for a normal modal logic $L$ with the universal
modality is to decide, given a formula $\varphi$, whether there exists a
substitution $\s$ such that $\s(\varphi)\in L$.
\end{definition}

\begin{theorem}\label{main1}
The unification problem for any normal modal logic between ${\ml K}_{u}$ and ${\ml K4}_u$ is undecidable.
\end{theorem}

The proof proceeds by reduction of some undecidable configuration problem for
Minsky machines.

We remind the reader that a {\em Minsky machine} (or a register machine with two
registers; see, e.g., \cite{Minsky61,Ebbinghaus1994}) is a finite set (program)
of instructions for transforming triples $\left\langle s,m,n\right\rangle$ of
natural numbers, called {\em configurations}. The intended meaning of the
current configuration $\left\langle s,m,n\right\rangle$ is as follows: $s$ is
the number (label) of the current machine state and $m$, $n$ represent the
current state of information. Each instruction has one of the four possible
forms:
\begin{align*}
& s\rightarrow\left\langle t,1,0\right\rangle, & & s\rightarrow\left\langle
t,-1,0\right\rangle( \left\langle t',0,0\right\rangle),\\
& s\rightarrow\left\langle t,0,1\right\rangle ,&& s\rightarrow\left\langle
t,0,-1\right\rangle( \left\langle t',0,0\right\rangle).
\end{align*}
The last of them, for instance, means: transform $\left\langle
s,m,n\right\rangle$ into $\left\langle t,m,n-1\right\rangle$ if $n>0$ and into
$\left\langle t',m,n\right\rangle$ if $n=0$. We assume that Minsky machines are
\emph{deterministic}, that is, they can have at most one instruction with a
given $s$ in the left-hand side. For a Minsky machine $\bp$, we write $\bp
:\left\langle s,m,n\right\rangle\rightarrow \left\langle t,k,l\right\rangle$ if
starting with $\left\langle s,m,n\right\rangle$ and applying the instructions in
$\bp$, in finitely many steps (possibly, in 0 steps) we can reach $\left\langle
t,k,l\right\rangle$.

We will use the well known fact (see, e.g., \cite{Chagrov90a,Chagrov&Z97}) that
there exist a Minsky program $\bp$ and a configuration ${\mathfrak
a}=\left\langle s,m,n\right\rangle$ such that no algorithm can decide, given a
configuration ${\mathfrak b}$, whether $\bp:{\mathfrak a}\rightarrow{\mathfrak
b}$.

Fix such a pair $\bp$ and $\mathfrak a=\left\langle s,m,n\right\rangle$, and
consider the transitive frame ${\mathfrak F}=(W,R)$ shown in Fig.~\ref{F5.2.1},
where the points $e(t,k,l)$ represent configurations $\left\langle
t,k,l\right\rangle$ such that $\bp :\left\langle s,m,n\right\rangle\rightarrow
\left\langle t,k,l\right\rangle$, $e(t,k,l)$ `sees' the points $a^{0}_t$,
$a^{1}_k$, $a^{2}_l$ representing the components of $\left\langle
t,k,l\right\rangle$, and $a$ is the only reflexive point of $\mathfrak F$. More
precisely,
\begin{multline*}
W ~=~  \{ a,b,g,g_1,g_2,d,d_1,d_2 \} \cup \{ a^i_j \mid i\le 2,\ j < \omega \}
\cup {} \\
  \{ e(t,k,l) \mid \bp :\left\langle s,m,n\right\rangle\rightarrow \left\langle
t,k,l\right\rangle \}
\end{multline*}
and $R$ is the transitive closure of the following relation:
\begin{multline*}
\{ (a,a), (g,a), (g,b), (d,b), (g_1,g), (g_2,g_1), (d_1,d), (d_2,d_1),\\
(a^0_0,g), (a^0_0,d), (a^1_0,g_1), (a^1_0,d_1), (a^2_0,g_2), (a^2_0,d_2)\} \cup
{} \phantom{MMMMMMM} \\
\{ (a^i_{j+1},a^i_j) \mid i\le 2,\ j<\omega \} \cup {} \phantom{MMMMMMMMM} \\
\{ \big( e(t,k,l), a^0_t \big ), \big( e(t,k,l), a^1_k \big ), \big( e(t,k,l),
a^2_l \big ) \mid e(t,k,l)\in W \}.
\end{multline*}
This frame and the formulas below describing it were introduced by A.~Chagrov in
\cite{Chagrov&Z97,Z&W&C} where the reader can find further references.

\begin{figure}
\begin{center}
\begin{picture}(170,180)(40,0)

\put(80,175){$\bullet$} \put(120,165){$\bullet$} \put(160,155){$\bullet$}
\put(200,145){$\bullet$} \put(40,155){$\circ$} \put(80,145){$\bullet$}
\put(120,135){$\bullet$} \put(160,125){$\bullet$} \put(100,125){$\bullet$}
\put(140,115){$\bullet$} \put(180,105){$\bullet$} \put(100,90){$\bullet$}
\put(107,90){$a_{1}^{0}$}

\put(140,90){$\bullet$} \put(147,90){$a_{1}^{1}$}

\put(180,90){$\bullet$} \put(187,90){$a_{1}^{2}$}

\put(100,75){$\bullet$} \put(107,75){$a_{2}^{0}$}

\put(140,75){$\bullet$} \put(147,75){$a_{2}^{1}$}

\put(180,75){$\bullet$} \put(187,75){$a_{2}^{2}$}

\put(101,63){\vdots} \put(141,63){\vdots} \put(181,63){\vdots}

\put(100,55){$\bullet$} \put(107,55){$a_{t-1}^{0}$}

\put(140,55){$\bullet$} \put(147,55){$a_{k-1}^{1}$}

\put(180,55){$\bullet$} \put(187,55){$a_{l-1}^{2}$}

\put(100,40){$\bullet$} \put(107,40){$a_{t}^{0}$}

\put(140,40){$\bullet$} \put(147,40){$a_{k}^{1}$}

\put(180,40){$\bullet$} \put(187,40){$a_{l}^{2}$}

\put(101,28){$\vdots$} \put(141,28){$\vdots$} \put(181,28){$\vdots$}

\put(120,10){$\bullet$} \put(117,1){$e(t,k,l)$}

\put(122,168){\vector(-4,1){37}} \put(162,158){\vector(-4,1){37}}
\put(202,148){\vector(-4,1){37}} \put(82,148){\vector(-4,1){37}}
\put(122,138){\vector(-4,1){37}} \put(162,128){\vector(-4,1){37}}
\put(102,129){\vector(-1,1){17}} \put(142,119){\vector(-1,1){17}}
\put(182,109){\vector(-1,1){17}} \put(102,127){\vector(1,2){20}}
\put(142,117){\vector(1,2){20}} \put(182,107){\vector(1,2){20}}

\put(106,12){\dots} \put(126,12){\dots}

\put(82,147){\vector(0,1){28}} \put(102,93.5){\vector(0,1){32}}
\put(142,93.5){\vector(0,1){22}} \put(182,93.5){\vector(0,1){12}}
\put(102,79){\vector(0,1){12}} \put(142,79){\vector(0,1){12}}
\put(182,79){\vector(0,1){12}}

\put(40,145){$a$} \put(80,135){$g$} \put(70,175){$b$} \put(120,175){$d$}
\put(160,165){$d_1$} \put(200,155){$d_2$} \put(120,145){$g_1$}
\put(160,135){$g_2$} \put(107,125){$a^{0}_{0}$} \put(147,115){$a_{0}^{1}$}
\put(187,105){$a_{0}^{2}$}

\put(102,44){\vector(0,1){12}} \put(142,44){\vector(0,1){12}}
\put(182,44){\vector(0,1){12}}

\put(122,13){\vector(-2,3){19}} \put(122.5,13){\vector(2,3){19}}
\put(123,13){\vector(2,1){58}}
\end{picture}
\end{center}
\caption{Frame $\mathfrak F$ encoding $\bp$ and $\mathfrak a$.} \label{F5.2.1}
\end{figure}

The following variable free formulas characterise the points in ${\mathfrak F}$
in the sense that each of these formulas, denoted by Greek letters with
subscripts and/or superscripts, is true in ${\mathfrak F}$ precisely at the
point denoted by the corresponding Roman letter with the same subscript and/or
superscript (and nowhere else):
\begin{align*}
&\alpha~=~\Diamond\top\wedge\Box\Diamond\top,\hspace*{3cm} \beta~=~\Box\bot,\\
&\gamma ~=~ \Diamond \alpha\wedge \Diamond
\beta\wedge\neg\Diamond^2\beta,\hspace*{2.1cm}
\delta ~=~\neg\gamma\wedge\Diamond\beta\wedge\neg\Diamond^2\beta,\\
&\delta_{1}~=~\Diamond \delta\wedge \neg\Diamond^2\delta,\hspace*{3cm}
\delta_{2}~=~\Diamond
\delta_{1}\wedge \neg\Diamond^2\delta_{1},\\
&\gamma_1~=~\Diamond\gamma\wedge\neg\Diamond^2\gamma\wedge\neg\Diamond\delta,\qquad
\hspace*{1.1cm}\gamma_2~=~\Diamond\gamma_1\wedge\neg\Diamond^2\gamma_1\wedge\neg\Diamond
\delta,\\
&\alpha_{0}^{0} ~=~\Diamond\gamma\wedge\Diamond\delta\wedge\neg\Diamond^2\gamma
\wedge\neg\Diamond^2\delta,\\
&\alpha_{0}^{1} ~=~\Diamond\gamma_{1}\wedge\Diamond\delta_{1}\wedge\neg\Diamond
^2\gamma_{1}\wedge\neg\Diamond^2\delta_{1},\\
&\alpha_{0}^{2} ~=~\Diamond\gamma_{2}\wedge\Diamond\delta_{2}\wedge\neg\Diamond
^2\gamma_{2}\wedge\neg\Diamond^2\delta_{2},\\
&\alpha_{j+1}^{i} ~=~ \Diamond \alpha_{0}^{i} \wedge
\Diamond\alpha_{j}^{i}\wedge\neg\Diamond^{2}\alpha_{j}^{i}
\wedge\bigwedge_{i\neq k}\neg\Diamond \alpha^k_0,
\end{align*}
where $i\in \{ 0,1,2\}$, $j\ge 0$.
It is worth emphasising that the formulas
\begin{equation}\label{property}
\alpha_{j}^{i} \rightarrow \neg \Diamond \alpha^{i}_{j} \quad \text{and} \quad
\alpha_{j+1}^{i} \rightarrow \Diamond \alpha^{i}_{0} \wedge \bigwedge_{k\not=i}
\neg\Diamond \alpha^{k}_{0}
\end{equation}
are valid in \emph{all frames} for all $i\in \{ 0,1,2\}$, $j\ge 0$. We will use
this property in what follows.

The formulas characterising the points $e(t,k,l)$ are denoted by
$\varepsilon(t,\alpha^{1}_{k},\alpha^{2}_{l})$ and defined as follows, where
$\varphi$ and $\psi$ are arbitrary formulas,
$$
\varepsilon(t,\varphi,\psi ) ~=~ \Diamond \alpha_{t}^{0}\wedge\neg\Diamond
\alpha_{t+1}^{0}\wedge\Diamond \varphi\wedge\neg\Diamond^2\varphi\wedge\Diamond
\psi\wedge\neg\Diamond^2\psi.
$$
We also require formulas characterising not only fixed but arbitrary
configurations:
\begin{align*}
\pi_{1} &~=~(\Diamond \alpha_{0}^{1}\vee \alpha_{0}^{1})\wedge\neg\Diamond
\alpha_{0}^{0}\wedge\neg\Diamond \alpha_{0}^{2}\wedge p_{1}\wedge\neg\Diamond
p_1,\\
\pi_{2} &~=~\Diamond \alpha_{0}^{1}\wedge\neg\Diamond
\alpha_{0}^{0}\wedge\neg\Diamond \alpha_{0}^{2}\wedge\Diamond
p_{1}\wedge\neg\Diamond^2p_{1},\\
\tau_{1} &~=~(\Diamond \alpha_{0}^{2}\vee \alpha_{0}^{2})\wedge\neg\Diamond
\alpha_{0}^{0}\wedge \neg\Diamond \alpha_{0}^{1}\wedge p_{2}\wedge\neg\Diamond
p_{2},\\
\tau_{2} &~=~\Diamond \alpha_{0}^{2}\wedge\neg\Diamond
\alpha_{0}^{0}\wedge\neg\Diamond \alpha_{0}^{1}\wedge\Diamond
p_{2}\wedge\neg\Diamond^2p_{2}.
\end{align*}
Observe that in $\mathfrak{F}$, under any valuation, $\pi_{1}$ can be true in at
most one point, and this point has to be $a^{1}_{j}$, for some $j\geq 0$.
Similarly, $\pi_{2}$ can only we true in at most one point, and this point has
to be of the form $a^{1}_{j}$, for some $j>0$. The same applies to $\tau_{1}$
and $\tau_{2}$, but with $a^{1}_{j}$ replaced by $a^{2}_{j}$.

Now we are fully equipped to simulate the behaviour of $\bp$ on $\mathfrak a$ by
means of modal formulas with the universal modalities.

With each instruction $I$ in $\bp$ we associate a formula $AxI$ by taking:
$$
AxI~=~\exists \varepsilon(t, \pi_{1},\tau_{1})\to \exists
\varepsilon(t',\pi_{2},\tau_{1})
$$
if $I$ is of the form $t\rightarrow\left\langle t',1,0\right\rangle$,
$$
AxI ~=~ \exists\varepsilon(t,\pi_{1},\tau_{1})\rightarrow
\exists\varepsilon(t',\pi_{1},\tau_{2})
$$
if $I$ is $t\rightarrow\left\langle t',0,1\right\rangle$,
\begin{eqnarray*}
AxI ~=~ \big(\exists\varepsilon(t,\pi_{2},\tau_{1})\rightarrow
\exists\varepsilon(t',\pi_{1},\tau_{1})\big)\wedge
\big(\exists\varepsilon(t,\alpha^{1}_{0},\tau_{1}) \rightarrow
\exists\varepsilon(t'',\alpha^{1}_{0},\tau_{1})\big)
\end{eqnarray*}
if $I$ is $t\rightarrow\left\langle t',-1,0\right\rangle (\left\langle
t'',0,0\right\rangle)$, and finally
\begin{eqnarray*}
AxI ~=~ \big(\exists\varepsilon(t,\pi_{1},\tau_{2})\rightarrow
\exists\varepsilon(t',\pi_{1},\tau_{1})\big)\wedge
\big(\exists\varepsilon(t,\pi_{1},\alpha^{2}_{0})\rightarrow
\exists\varepsilon(t'',\pi_{1},\alpha^{2}_{0})\big)
\end{eqnarray*}
if $I$ is $t\rightarrow\left\langle t',0,-1\right\rangle (\left\langle
t'',0,0\right\rangle)$.

The formula simulating $\bp$ as a whole is
$$
AxP ~=~ \bigwedge_{I\in\mbox{\scriptsize $\bp$}}AxI.
$$
One can readily check that $\mathfrak F \models AxP$.

Now, for each $\mathfrak b = \langle t,k,l\rangle$  consider the formula
$$
\psi(\mathfrak b) ~=~ \big ( AxP \land \exists\varepsilon
(s,\alpha^1_m,\alpha^2_n) \big ) \to \exists\varepsilon
(t,\alpha^1_k,\alpha^2_l).
$$
\begin{lemma}\label{main-lemma}
Let ${\ml K}_{u} \subseteq L \subseteq {\ml K4}_u$. Then $\bp:{\mathfrak
a}\rightarrow{\mathfrak b}$ iff $\psi(\mathfrak b)$ is unifiable in $L$.
\end{lemma}
\begin{proof}[of Lemma]
$(\Leftarrow)$ Suppose that $\bp:{\mathfrak a} \not\to{\mathfrak b}$. Then, by
the construction of $\mathfrak F$, we have
$$
\mathfrak F \models AxP \land \exists\varepsilon (s,\alpha^1_m,\alpha^2_n) \quad
\text{and}\quad \mathfrak F \not\models \exists\varepsilon
(t,\alpha^1_k,\alpha^2_l).
$$
As $\exists\varepsilon (t,\alpha^1_k,\alpha^2_l)$ is variable free, all
substitution instances of $\psi(\mathfrak b)$ are refuted in $\mathfrak F$, and
so $\psi(\mathfrak b)$ is not unifiable in any $L\subseteq {\ml K4}_u$.

\smallskip

$(\Rightarrow)$ Conversely, suppose that $\bp:{\mathfrak a} \to{\mathfrak b}$.
Our aim is to find a substitution $\s$ for the variables $p_1$ and $p_2$ such
that $\s (\psi(\mathfrak b)) \in {\ml K}_u$.

Let
$$
\bp : \mathfrak a = \langle t_0,k_0,l_0\rangle \stackrel{I_1}\to \langle
t_1,k_1,l_1\rangle \stackrel{I_2}\to \dots \stackrel{I_\ell}\to \langle
t_\ell,k_\ell,l_\ell\rangle = \mathfrak b
$$
be the computation of $\bp$ starting with $\mathfrak a$ and ending with
$\mathfrak b$, where $I_j$ is the instruction from $\bp$ that is used to
transform $\langle t_{j-1},k_{j-1},l_{j-1}\rangle$ into $\langle
t_j,k_j,l_j\rangle$. Consider the formula
\begin{equation}\label{defect}
{\sf defect}_{i} ~=~  \exists \varepsilon (t_0, \alpha^1_{k_0}, \alpha^2_{l_0})
\land \dots \land \exists \varepsilon (t_i, \alpha^1_{k_i}, \alpha^2_{l_i})
\land \neg \exists \varepsilon (t_{i+1}, \alpha^1_{k_{i+1}}, \alpha^2_{l_{i+1}})
\end{equation}
which `says' that the computation is simulated properly up to the $i$th step,
but there is no point representing the $i+1$st configuration.

Define the substitution $\s$ we need by taking
\begin{equation}\label{substitution}
\s (p_1) ~=~ \bigvee_{i=0}^{\ell-1} {\sf defect}_{i} \land
\overline{\alpha}^1_{k_i}, \qquad  \s (p_2) ~=~ \bigvee_{i=0}^{\ell-1} {\sf
defect}_{i} \land \overline{\alpha}^2_{l_i},
\end{equation}
where
$$
\overline{\alpha}^{1}_{k_{i}} ~=~
\begin{cases}
\alpha^{1}_{k_{i}} & \text{if either}\ k_{i} = 0\ \text{or}\ I_{i+1} \ne t_i \to \langle t_{i+1},-1,0\rangle,\\
\alpha^{1}_{k_{i}-1} & \text{if} \ k_{i} \ne 0 \ \text{and}\ I_{i+1} = t_i \to
\langle t_{i+1},-1,0\rangle,
\end{cases}
$$
and
$$
\overline{\alpha}^{2}_{l_{i}} ~=~
\begin{cases}
\alpha^{2}_{l_{i}} & \text{if either}\ l_{i} = 0\ \text{or}\ I_{i+1} \ne t_i \to \langle t_{i+1},0,-1\rangle,\\
\alpha^{2}_{l_{i}-1} & \text{if} \ l_{i} \ne 0 \ \text{and}\  I_{i+1} = t_i \to
\langle t_{i+1},0,-1\rangle.
\end{cases}
$$
We show now that we have $\mathfrak G \models \s (\psi(\mathfrak b))$ for
\emph{all} frames $\mathfrak G$, which clearly means that $\s (\psi ({\mathfrak
b})) \in {\ml K}_u$.

Suppose $\mathfrak G = (W,R)$ is given. As all formulas considered below, in
particular $\s (\psi ({\mathfrak b}))$, are variable free, we can write $x
\models \psi$ to say that $\psi$ is true at $x$ in some/all models based on
${\mathfrak G}$. Moreover, for any Boolean combination $\psi$ of such formulas
starting with $\exists$, we have $x\models \psi$ iff $x'\models \psi$ for any
$x,x'\in W$. Hence, ${\mathfrak G}\not\models \psi$ means that
$x\not\models\psi$ for all $x\in W$.

Let us now proceed with the proof. Two cases are possible.

\smallskip

\emph{Case} 1: $\mathfrak G \models \neg \exists\varepsilon
(t_0,\alpha^1_{k_0},\alpha^2_{l_0}) \lor \exists\varepsilon
(t_\ell,\alpha^1_{k_\ell},\alpha^2_{l_\ell})$. Then clearly $\mathfrak G \models
\s (\psi(\mathfrak b))$.

\smallskip

\emph{Case} 2: $\mathfrak G \models \exists\varepsilon
(t_0,\alpha^1_{k_0},\alpha^2_{l_0}) \land \neg \exists\varepsilon
(t_\ell,\alpha^1_{k_\ell},\alpha^2_{l_\ell})$. Then there exists some number $i
< \ell$ such that $\mathfrak G \models {\sf defect}_{i}$. It follows that, for
all $z\in W$,
\begin{equation}\label{property1}
z \models \s(p_{1}) \quad \text{iff}\quad z\models
\overline{\alpha}_{k_{i}}^{1}, \quad \text{and} \quad z\models \s(p_{2}) \quad
\text{iff} \quad z\models \overline{\alpha}_{l_{i}}^{2}.
\end{equation}

\begin{claim}\label{claim1}
For all $z\in W$, we have {\rm (i)} $z \models \s(\pi_{1})$ iff
$z \models \overline{\alpha}_{k_{i}}^{1}$, and {\rm (ii)} $z \models
\s(\tau_{1})$ iff $z \models \overline{\alpha}_{l_{i}}^{2}$.
\end{claim}
\begin{proof}[of Claim]
Suppose $z\in W$ is given. We know that
$$
\s(\pi_{1}) ~=~ (\Diamond \alpha_{0}^{1} \vee \alpha_{0}^{1}) \wedge \neg
\Diamond \alpha_{0}^{0} \wedge \neg \Diamond \alpha_{0}^{2} \wedge \s(p_{1})
\wedge \neg \Diamond \s(p_{1}).
$$
Hence, by \eqref{property1} and \eqref{property},
$$
z \models \s(\pi_{1}) \quad \text{iff} \quad z \models (\Diamond \alpha_{0}^{1}
\vee \alpha_{0}^{1}) \wedge \neg \Diamond \alpha_{0}^{0} \wedge \neg \Diamond
\alpha_{0}^{2} \wedge \overline{\alpha}_{k_{i}}^{1} \wedge \neg \Diamond
\overline{\alpha}_{k_{i}}^{1} \quad \text{iff} \quad z \models
\overline{\alpha}_{k_{i}}^{1}.
$$
(ii) is considered analogously.
\end{proof}

\begin{claim}\label{claim2}
For all $z \in W$, {\rm (i)} $z \models \s(\pi_{2})$ iff $z \models
\overline{\alpha}_{k_{i}+1}^{1}$, and {\rm (ii)} $z \models \s(\tau_{2})$ iff $z
\models \overline{\alpha}_{l_{i}+1}^{2}$.
\end{claim}
\begin{proof}[of Claim]
Suppose $z\in W$ is given. We know that
$$
\s(\pi_{2}) ~=~ \Diamond \alpha_{0}^{1} \wedge \neg \Diamond \alpha_{0}^{0}
\wedge \neg \Diamond \alpha_{0}^{2} \wedge \Diamond \s(p_{1}) \wedge \neg
\Diamond^{2} \s(p_{1}).
$$
Hence, by \eqref{property1},
$$
z \models \s(\pi_{2}) \quad \text{iff} \quad z \models \Diamond \alpha_{0}^{1}
\wedge \neg \Diamond \alpha_{0}^{0} \wedge \neg \Diamond \alpha_{0}^{2} \wedge
\Diamond \overline{\alpha}_{k_{i}}^{1} \wedge \neg \Diamond^{2}
\overline{\alpha}_{k_{i}}^{1}.
$$
But, according to \eqref{property}, the latter formula is equivalent to the
definition of $\overline{\alpha}^{1}_{k_{i}+1}$, which proves the claim.
\end{proof}

We now make a case distinction according to rule $I_{i+1}$ used to transform
$\langle t_i,k_i,l_i\rangle$ to $\langle t_{i+1},k_{i+1},l_{i+1}\rangle$.

\smallskip

\emph{Case} 1: $I_{i+1} = t_i \to \langle t_{i+1},1,0\rangle$. Our aim is to
show that
\begin{itemize}
\item[(a)] ${\mathfrak G} \models \s (\exists \varepsilon(t_{i},\pi_{1},\tau_{1}))$
and

\item[(b)] ${\mathfrak G} \not\models \s (\exists \varepsilon(t_{i+1},\pi_{2},
\tau_{1}))$,
\end{itemize}
for then we would have ${\mathfrak G}\not\models \s(AxP)$, and so ${\mathfrak
G}\models \s(\psi({\mathfrak b}))$.

\smallskip

(a) As ${\mathfrak G} \models \exists
\varepsilon(t_{i},\alpha_{k_{i}}^{1},\alpha_{l_{i}}^{2})$, we have some $z\in W$
such that
$$
z \models \Diamond \alpha_{t_{i}}^{0}\wedge\neg\Diamond
\alpha_{t_{i}+1}^{0}\wedge\Diamond \alpha_{k_{i}}^{1} \wedge\neg\Diamond^2\alpha_{k_{i}}^{1}\wedge
\Diamond \alpha_{l_{i}}^{2}\wedge\neg\Diamond^2\alpha_{l_{i}}^{2}.
$$
By Claim~\ref{claim1}, we then have
$$
z \models \Diamond \alpha_{t_{i}}^{0}\wedge\neg\Diamond \alpha_{t_{i}+1}^{0}
\wedge \Diamond \s(\pi_{1}) \wedge \neg \Diamond^{2} \s(\pi_{1}) \wedge \Diamond
\s(\tau_{1}) \wedge \neg \Diamond^{2} \s (\tau_{1}),
$$
which means that $z \models \s (\varepsilon(t_{i},\pi_{1},\tau_{1}))$, and so
${\mathfrak G} \models \s (\exists \varepsilon(t_{i},\pi_{1},\tau_{1}))$.

\smallskip

(b) Suppose that $\mathfrak G \not\models \s(\exists \varepsilon(t_{i+1},
\pi_{2},\tau_{1}))$ does not hold. Then there is $x\in W$ with
$$
x \models \varepsilon(t_{i+1}, \s(\pi_{2}), \s(\tau_{1})),
$$
that is,
$$
x \models \Diamond \alpha_{t_{i+1}}^{0}\wedge\neg\Diamond
\alpha_{t_{i+1}+1}^{0}\wedge\Diamond \s(\pi_2) \wedge\neg\Diamond^2 \s(\pi_2)
\wedge \Diamond \s(\tau_1) \wedge\neg \Diamond^2\s(\tau_1).
$$
By Claims~\ref{claim1} and \ref{claim2}, we then have
$$
x \models \Diamond \alpha_{t_{i+1}}^{0}\wedge\neg\Diamond
\alpha_{t_{i+1}+1}^{0}\wedge\Diamond \alpha_{k_{i}+1}^{1} \wedge\neg\Diamond^2 \alpha_{k_{i}+1}^{1}
\wedge \Diamond \alpha_{l_{i}}^{2} \wedge\neg \Diamond^2 \alpha_{l_{i}}^{2}
$$
which means
$$
x \models \varepsilon(t_{i+1}, \alpha_{k_{i}+1}^{1}, \alpha_{l_{i}}^{2}).
$$
Now recall that $\alpha_{k_{i}+1}^{1}= \alpha_{k_{i+1}}^{1}$ and
$\alpha_{l_{i}}= \alpha_{l_{i+1}}$, that is, we have
$$
x \models \varepsilon(t_{i+1}, \alpha_{k_{i+1}}^{1}, \alpha_{l_{i+1}}^{2}),
$$
and so ${\mathfrak G} \models \exists \varepsilon(t_{i+1}, \alpha_{k_{i+1}}^{1},
\alpha_{l_{i+1}}^{2})$, contrary to ${\mathfrak G}\models {\sf defect}_{i}$.

\smallskip

\emph{Case} 2: $I_{i+1}$ is of the form $t_i \to \langle t'_{i+1},-1,0\rangle
(\langle t''_{i+1},0,0 \rangle)$. Suppose first that $k_i=0$, that is, the
actual instruction is $I_{i+1}= t_i \to \langle t_{i+1},0,0 \rangle$. We need to
show that
\begin{itemize}
\item[(a)] ${\mathfrak G} \models \s (\exists \varepsilon(t_{i},\alpha_{0}^{1},
\tau_{1}))$ and

\item[(b)] ${\mathfrak G} \not\models \s (\exists
\varepsilon(t_{i+1},\alpha_{0}^{1},\tau_{1}))$,
\end{itemize}
which, as before, would imply ${\mathfrak G}\models \s(\psi({\mathfrak b}))$.

\medskip

(a) As ${\mathfrak G} \models \exists
\varepsilon(t_{i},\alpha_{0}^{1},\alpha_{l_{i}}^{2})$, we have $x\in W$ such
that
$$
x \models \Diamond \alpha_{t_{i}}^{0}\wedge\neg\Diamond
\alpha_{t_{i}+1}^{0}\wedge\Diamond \alpha_{0}^{1}
\wedge\neg\Diamond^2\alpha_{0}^{1}\wedge \Diamond
\alpha_{l_{i}}^{2}\wedge\neg\Diamond^2\alpha_{l_{i}}^{2},
$$
from which, by Claim~\ref{claim1},
$$
x \models \Diamond \alpha_{t_{i}}^{0}\wedge\neg\Diamond \alpha_{t_{i}+1}^{0}
\wedge \Diamond \alpha_{0}^{1} \wedge \neg \Diamond^{2} \alpha_{0}^{1} \wedge
\Diamond \s(\tau_{1}) \wedge \neg \Diamond^{2} \s (\tau_{1}).
$$
Thus we have $x \models \exists \varepsilon(t_{i},\alpha_{0}^{1}, \tau_{1})$.
(b) is proved similarly and left to the reader.

\smallskip

Suppose now that $k_{i}>0$, that is, the instruction $I_{i+1}= t_i \to \langle
t_{i+1},-1,0\rangle$ was actually used. This time we need to show that
\begin{itemize}
\item[(a)] ${\mathfrak G} \models \s(\exists
\varepsilon(t_{i},\pi_{2},\tau_{1}))$ and

\item [(b)] ${\mathfrak G} \not\models \s (\exists \varepsilon(t_{i+1},\pi_{1},
\tau_{1}))$.
\end{itemize}

(a) Since ${\mathfrak G} \models \exists
\varepsilon(t_{i},\alpha_{k_{i}}^{1},\alpha_{l_{i}}^{2})$, we have $x\in W$ such
that
$$
x \models \Diamond \alpha_{t_{i}}^{0}\wedge\neg\Diamond
\alpha_{t_{i}+1}^{0}\wedge\Diamond \alpha_{k_{i}}^{1} \wedge\neg\Diamond^2\alpha_{k_{i}}^{1}\wedge
\Diamond \alpha_{l_{i}}^{2}\wedge\neg\Diamond^2\alpha_{l_{i}}^{2}.
$$
Clearly, it is sufficient to show that
$$
x \models \Diamond \alpha_{t_{i}}^{0}\wedge\neg\Diamond \alpha_{t_{i}+1}^{0}
\wedge \Diamond \s (\pi_{2}) \wedge \neg \Diamond^{2} \s (\pi_{2}) \wedge
\Diamond \s (\tau_{1}) \wedge \neg \Diamond^{2} \s (\tau_{1}).
$$
Observe that in this case $\overline{\alpha}_{k_{i}}^{1}= \alpha_{k_{i}-1}$.
Hence, by Claim~\ref{claim2}, for all $z\in W$ we have $z \models \s (\pi_{2})$
iff $z \models \alpha^1_{k_{i}}$. So it remains to use Claims~\ref{claim1} and
\ref{claim2}.

\smallskip

(b) Suppose otherwise, that is, ${\mathfrak G}\models \s
(\exists\varepsilon(t_{i+1},\pi_{1},\tau_{1}))$. Then there exists $x \in W$
such that
$$
x \models \Diamond \alpha_{t_{i+1}}^{0}\wedge\neg\Diamond
\alpha_{t_{i+1}+1}^{0}\wedge\Diamond \s(\pi_1) \wedge\neg\Diamond^2 \s(\pi_1)
\wedge \Diamond \s(\tau_1) \wedge\neg \Diamond^2\s(\tau_1).
$$
By Claim~\ref{claim1}, this implies
$$
x \models \Diamond \alpha_{t_{i+1}}^{0}\wedge\neg\Diamond
\alpha_{t_{i+1}+1}^{0}\wedge\Diamond \alpha_{k_{i}-1}^{1} \wedge\neg\Diamond^2
\alpha_{k_{i}-1}^{1} \wedge \Diamond \alpha_{l_{i}}^{2} \wedge\neg \Diamond^2
\alpha_{l_{i}}^{2},
$$
that is,
$$
x \models \varepsilon(t_{i+1}, \alpha_{k_{i}-1}^{1}, \alpha_{l_{i}}^{2})
$$
which leads to a contradiction, because  $\alpha_{k_{i}-1}^{1}=
\alpha_{k_{i+1}}^{1}$ and $\alpha_{l_{i}} = \alpha_{l_{i+1}}$, and therefore we
must have ${\mathfrak G} \models \exists \varepsilon(t_{i+1},
\alpha_{k_{i+1}}^{1}, \alpha_{l_{i+1}}^{2})$.

\smallskip

The remaining two types of instructions (where the third component changes) are
dual to the ones considered above. We leave these cases to the reader.

This completes the proof of Lemma~\ref{main-lemma}. Theorem~\ref{main1} follows
immediately in view of the choice of $\bp$ and $\alpha$.
\end{proof}

Observe that Theorem~\ref{main1} can be proved for multimodal ${\sf K}_{u}$ and
${\sf K4}_{u}$ as well. In this case, in the frame ${\mathfrak F}$ considered
above, the additional operators can be interpreted by the empty relation. By a
proper modification of the frame $\mathfrak F$ in Fig.~\ref{F5.2.1}, this
theorem can also be extended to some logics above ${\sf K4}_{u}$, for example,
${\sf GL}_{u}$.

\begin{definition} The \emph{admissibility problem} for inference rules for a
normal modal logic $L$ with the universal modality is to decide, given an
inference rule $\varphi_1,\dots,\varphi_n /\varphi$, whether $\s(\varphi_1)\in
L$, \dots, $\s(\varphi_n)\in L$ imply $\s(\varphi)\in L$, for every substitution
$\s$.
\end{definition}

As an immediate consequence of Theorem~\ref{main1} we obtain the following:

\begin{theorem}
The admissibility problem for any normal modal logic $L$ between ${\ml K}_{u}$
and ${\ml K4}_{u}$ is undecidable.
\end{theorem}

Minor modifications of the proof above can be used to prove undecidability of
the unification and admissibility problems for various modal logics in which the
universal modality is definable. An interesting example is {\ml PDL} with
converse, i.e., the extension of propositional dynamic logic with the converse
constructor on programs: if $\alpha$ is a program, then $\alpha^{-1}$ is a
program which is interpreted by the converse of the relation interpreting
$\alpha$. (We do not provide detailed definitions of the syntax and semantics
here but refer the reader to \cite{Hareletal00}.) The undecidability proof for
the unification problem (for substitutions instead of propositional variables
rather than atomic programs!) is carried out by taking an atomic program
$\alpha$ and replacing, in the proof above, the operator $\Box$ with $[\alpha]$
and the universal modality $\forall$ with $[(\alpha \cup \alpha^{-1})^{\ast}]$.

It seems worth mentioning, however, that the unification problem is trivially
decidable for any normal modal logic $L$ with $\neg \Box \bot\in L$. To see
this, recall that a substitution $\s$ is called \emph{ground} if it replaces
each propositional variable by a variable free formula (that is, a formula
constructed from $\bot$ and $\top$ only). Obviously, it is always the case that
if there exists a substitution $\s$ such that $\s(\varphi)\in L$, then there
exists a ground substitution $\s'$ with $\s'(\varphi)\in L$. But if $\neg \Box
\bot \in L$, then there are, up to equivalence in $L$, only two different
variable free formulas, namely, $\bot$ and $\top$. Thus, to decide whether a
formula $\varphi$ is unifiable in $L$ it is sufficient to check whether any of
the ground substitutions makes $\varphi$ equivalent to $\top$ (which can be done
in Boolean logic). A well known example of such a logic is ${\ml S4}_{u}$, ${\ml
S4}$ with the universal modality. Note that the admissibility problem for ${\ml
S4}_{u}$ might nevertheless be undecidable. We leave this as an interesting open
problem.


\section{Unification modulo equational theories}

The results presented above can be reformulated as undecidability results for
the well-known notion of unification modulo equational theories
\cite{Baader&Siekmann,BaaderSnyderHandbook00}.

Consider the equational theory ${\sf BAO}_{2}$ of Boolean algebras with
operators $\Box_{1}$ and $\Box_{2}$, which consists of an axiomatisation {\sf
BA} of the variety of Boolean algebras (say, in the signature with the binary
connective $\wedge$, unary connective $\neg$ and constant $1$) together with the
equations
$$
\Box_{i}(x \wedge y) ~=~ \Box_{i}x \wedge \Box_{i} y \quad \text{and} \quad
\Box_{i} 1 ~=~ 1,
$$
for $i=1,2$. Let $T$ be any set of equations over the signature of Boolean
algebras with two operators. Then the \emph{unification problem modulo} ${\sf
BAO}_{2} \cup T$ is to decide, given an equation $t_{1}=t_{2}$ over the
signature of ${\sf BAO}_{2}$, whether there exists a substitution $\s$ such that
$$
\s(t_{1}) ~=_{{\sf BAO}_{2} \cup T}~ \s (t_{2}),
$$
that is, whether there exists a substitution $\s$ such that the equation
$\s(t_{1}) = \s (t_{2})$ is valid in all algebras where the equations in ${\sf
BAO}_{2} \cup T$ hold true. For a term $t$, let $t^{p}$ denote the propositional
modal formula that is obtained from $t$ by replacing its (individual) variables
with (mutually distinct) propositional variables. We may assume that $\cdot^{p}$
is a bijection between the terms $t$ over the signature of ${\sf BAO}_{2}$ and
the modal formulas with modal operators $\Box_{1}$ and $\Box_{2}$. Denote by
$\cdot^{-p}$ the inverse of this function. It is well-known (see, e.g.,
\cite{yde}) that a modal formula $\varphi$ is valid in the smallest normal modal
logic $L$ containing the formulas
$$
\{ t_{1}^{p} \leftrightarrow t_{2}^{p} \mid t_{1}=t_{2} \in T\}
$$
if, and only if, $\varphi^{-p}$ is valid in all algebras validating ${\sf
BAO}_{2} \cup T$. The appropriate converse statement is also easily formulated.
It follows that the unification problem modulo ${\sf BAO}_{2} \cup T$ is
decidable if, and only if, the unification problem for $L$ is decidable.
Clearly, it remains an open question whether the unification problem modulo
${\sf BAO}_{2}$ is decidable. However, if $T$ consists of the following
inequalities (saying that $\Box_{1}$ is the universal box)
$$
\Box_{1} x ~\leq~ \Box_{2}x, \quad \Box_{1} x ~\leq~ x, \quad \Box_{1} x ~\leq~
\Box_{1}\Box_{1} x, \quad x ~\leq~ \Box_{1}\neg\Box_{1}\neg x,
$$
then Theorem~\ref{main1} implies that the unification problem modulo ${\sf BAO}_{2} \cup T$ is undecidable.


\section{Unification in modal logics with nominals}

Let us now consider the extension of the language ${\cal L}$ with nominals. More
precisely, denote by $\mathcal{H}_{2}$ the propositional language constructed
from
\begin{itemize}
\item an infinite list $p_{1},p_{2},\dots$ of propositional variables and

\item an infinite list $n_{1},n_{2},\dots$ of \emph{nominals}
\end{itemize}
using the standard Boolean connectives and two modal operators $\Box$ and
$\Box_{h}$ (instead of $\Box$ and $\forall$ in ${\cal L}$).\footnote{The
language with infinitely many modal operators and nominals is often denoted by
$\mathcal{H}$ and called the \emph{minimal hybrid logic}; see, e.g.,
\cite{Arecesten}.} $\mathcal{H}_{2}$-formulas are interpreted in frames of the
form ${\mathfrak F}=(W,R,S)$ where $R,S \subseteq W\times W$. As before, a
\emph{model} is a pair $\mathfrak M = ({\mathfrak F},{\mathfrak V})$, where
${\mathfrak V}$ is a \emph{valuation function} that assigns to each $p_{i}$ a
subset $\mathfrak V(p_i)$ of $W$ and to each $n_{i}$ a \emph{singleton subset}
$\mathfrak V(n_i)$ of $W$. The \emph{truth-relation}, $(\mathfrak M, x) \models
\varphi$, is defined as above with two extra clauses:
\begin{itemize}
\item[] $(\mathfrak M,x) \models n_{i}$ iff $\{x\} = {\mathfrak V}(n_{i})$,

\item[] $(\mathfrak M,x) \models \Box_h\psi$ iff $(\mathfrak M,y) \models \psi$ for
all $y\in W$ with $xSy$.
\end{itemize}

Denote by ${\sf K}_{{\mathcal H}_{2}}$ the set of all $\mathcal{H}_{2}$-formulas
that are valid in all frames, and denote by ${\sf K}_{{\mathcal H}_{2}}\oplus
45$ the set of $\mathcal{H}_{2}$-formulas that are valid in all frames $(W,R,S)$
with transitive $R$ and $S= W \times W$. A proof of the following result can be
found in \cite{Arecesetal00}:

\begin{fact}
The satisfiability problem for ${\sf K}_{{\mathcal H}_{2}}$ is {\sc
PSpace}-complete, while for ${\sf K}_{{\mathcal H}_{2}}\oplus 45$ it is {\sc
ExpTime}-complete.
\end{fact}

A \emph{substitution} $\s$ for $\mathcal{H}_{2}$ is a map from the set of
propositional variables into $\mathcal{H}_{2}$. In particular, any substitution
leaves nominals intact.\footnote{Alternatively, we could allow nominals to be
substituted by nominals. This would not affect the undecidability result.} The
unification and admissibility problems for modal logics with nominals are
formulated in exactly the same way as before.

\begin{theorem}\label{main2}
The unification problem and, therefore, the admissibility problem for any logic
$L$ between ${\ml K}_{{\mathcal H}_{2}}$ and  ${\ml K}_{{\mathcal H}_{2}}\oplus
45$ are undecidable.
\end{theorem}

The proof of this theorem is similar to the proof of Theorem~\ref{main1}. Here
we only show how to modify the encoding of Minsky machine computations from
Section~\ref{universal}. The main difference is that now the language does not
contain the universal modality which can refer to all points in the frame in
order to say, e.g., that a certain configuration is (not) reachable. To overcome
this problem, we will use one nominal, let us call it $n$, which, if accessible
from a point $x$ (via $R$ and $S$), will be forced to be accessible from all
points located within a certain distance from $x$. This trick will provide us
with a `surrogate' universal modality which behaves, locally, similarly to the
standard one.

From now on we will be using the following abbreviation, where $\varphi$ is an
$\mathcal{H}_{2}$-formula:
\begin{equation}\label{surrogate}
\exists \varphi ~=~ \Diamond_{h}(n \wedge \Diamond_{h} \varphi).
\end{equation}
The defined operator $\exists$ will play the role of our surrogate universal
diamond.

Consider again a Minsky program $\bp$ and a configuration ${\mathfrak a}=
\left\langle s,m,n \right\rangle$ such that it is undecidable, given a
configuration $\mathfrak b$, whether $\bp : {\mathfrak a} \rightarrow {\mathfrak
b}$. The frame ${\mathfrak F} = (W,R,S)$ encoding $\mathfrak F$ and $\mathfrak
a$ is defined as in Fig.~\ref{F5.2.1}, with $S = W \times W$. For each
instruction $I$, we introduce the formula $AxI$ in precisely the same way as
before, with $\exists$ defined by \eqref{surrogate}.

The first important difference between the two constructions is the definition
of $AxP$. Let $\Ax$ denote the conjunction of all $\mathcal{H}_{2}$-formulas of
the form
$$
\Diamond_{h}n \rightarrow M \Diamond_{h} n \quad \mbox{and} \quad M'
\Diamond_{h}n \rightarrow \Diamond_{h} n,
$$
where $M$ is any sequence of $\Box$ and $\Box_{h}$ of length $\le 6$, and $M'$
is any sequence of $\Diamond$ and $\Diamond_{h}$ of length $\le 6$. To explain
the meaning of $\Ax$, consider a model $({\mathfrak G},{\mathfrak V})$ based on
some frame ${\mathfrak G}=(W,R,S)$. Let $x_{0}\in W$. We say that $x\in W$ is
\emph{of distance} $\le m$ \emph{from} $x_{0}$ if there exists a sequence
$$
x_{0} S' x_{1} S' x_{2} \cdots x_{k-1} S 'x_{k} ~ = ~ x,
$$
where $S' = R \cup S$ and $k\le m$. Now assume that $x_{0} \models \Ax$. Then
either all points of distance $\le 6$ from $x_{0}$ `see' ${\mathfrak V}(n)$ via
$S$, or no point of distance $\le 6$ from $x_{0}$ sees ${\mathfrak V}(n)$ via
$S$. In particular, $x_{0}\models \exists \varphi$ if, and only if, $x\models
\exists \varphi$ for all $x$ of distance $\le 6$ from $x_{0}$, and
$x_{0}\not\models\exists\varphi$ if, and only if, $x\not\models \exists\varphi$
for all $x$ of distance $\le 6$ from $x_{0}$.

The formula simulating $\bp$ as a whole in this case is
$$
AxP ~=~ \bigwedge_{I\in\mbox{\scriptsize $\bp$}}AxI \wedge \Ax.
$$

Consider the frame ${\mathfrak F}=(W,R,S)$ in Fig.~\ref{F5.2.1} (with $S=W\times
W$). Then, no matter which singleton set interprets $n$, the new operator
$\exists$ is always interpreted by the universal relation. Hence, as before we
have $\mathfrak F \models AxP$.

Now, for each $\mathfrak b = \langle t,k,l\rangle$ consider (as before) the formula
$$
\psi(\mathfrak b) ~=~ AxP \land \exists\varepsilon (s,\alpha^1_m,\alpha^2_n) \to
\exists\varepsilon (t,\alpha^1_k,\alpha^2_l).
$$
\begin{lemma}\label{lem3}
$\bp:{\mathfrak a}\rightarrow{\mathfrak b}$ iff $\psi(\mathfrak b)$ is unifiable
in $L$, where ${\ml K}_{{\mathcal H}_{2}} \subseteq L \subseteq {\sf
K}_{{\mathcal H}_{2}}\oplus 45$.
\end{lemma}
\begin{proof}[of Lemma]
The proof of $(\Leftarrow)$ is exactly as before.

\smallskip

$(\Rightarrow)$ Suppose that $\bp:{\mathfrak a} \to{\mathfrak b}$. Our aim is to
find a substitution $\s$ for the variables $p_1$ and $p_2$ such that $\s
(\psi(\mathfrak b)) \in {\ml K}_{{\mathcal H}_{2}}$. The definition of the
substitution is as before. Let
$$
\bp : \mathfrak a = \langle t_0,k_0,l_0\rangle \stackrel{I_1}\to \langle
t_1,k_1,l_1\rangle \stackrel{I_2}\to \dots \stackrel{I_\ell}\to \langle
t_\ell,k_\ell,l_\ell\rangle = \mathfrak b
$$
be the computation of $\bp$ starting with $\mathfrak a$ and ending with
$\mathfrak b$. Then we define $\s$ by means of \eqref{substitution}, where ${\sf
defect}_i$ is given by \eqref{defect}.

We have to show that, for \emph{all} frames $\mathfrak G$, we have $\mathfrak G
\models \s (\psi(\mathfrak b))$. Note that now we \emph{cannot assume} that
$\exists$ is interpreted by the universal relation.

Suppose that we are given a frame $\mathfrak G = (W,R,S)$, a valuation
${\mathfrak V}$ in it, and some $x_{0}\in W$. We write $\{n^{\mathfrak V}\}$ for
${\mathfrak V}(n)$, and $x\models \psi$ for $({\mathfrak G},{\mathfrak
V},x)\models \psi$. As before, two cases are possible.

\smallskip

\emph{Case} 1: $x_{0} \models \neg \exists\varepsilon
(t_0,\alpha^1_{k_0},\alpha^2_{l_0}) \lor \exists\varepsilon
(t_\ell,\alpha^1_{k_\ell},\alpha^2_{l_\ell})$. Then clearly $x_{0} \models
\s (\psi(\mathfrak b))$.

\smallskip

\emph{Case} 2: $x_{0} \models \exists\varepsilon
(t_0,\alpha^1_{k_0},\alpha^2_{l_0}) \land \neg \exists\varepsilon
(t_\ell,\alpha^1_{k_\ell},\alpha^2_{l_\ell})$. If $x_{0} \not\models \s(\Ax)$
then obviously $x_{0} \models \s(\psi({\mathfrak b}))$, and we are done. So
assume that $x_{0} \models \s(\Ax)$. Then there exists some number $i < \ell$
such that $x_{0} \models {\sf defect}_{i}$.

\begin{claim}\label{closer}
For all points $x$ of distance $\le 6$ from $x_{0}$, $x \models {\sf
defect}_{i}$. So, for all such $x$, we have $x \models \s(p_{1})$ iff $x\models
\overline{\alpha}_{k_{i}}^{1}$, and $x\models \s(p_{2})$ iff $x\models
\overline{\alpha}_{l_{i}}^{2}$.
\end{claim}
\begin{proof}[of Claim]
Follows immediately from $x_{0} \models \Ax$.
\end{proof}

\begin{claim}\label{again1}
For all $x$ of distance $\le 5$ from $x_{0}$, we have {\rm (i)} $x \models
\s(\pi_{1})$ iff $x \models \overline{\alpha}_{k_{i}}^{1}$, and {\rm (ii)} $x
\models \s(\tau_{1})$ iff $x \models \alpha_{l_{i}}^{2}$.
\end{claim}
\begin{proof}[of Claim]
We only prove (i). Suppose $x$ is given. We know that
$$
\s(\pi_{1}) ~=~ (\Diamond \alpha_{0}^{1} \vee \alpha_{0}^{1}) \wedge \neg
\Diamond \alpha_{0}^{0} \wedge \neg \Diamond \alpha_{0}^{2} \wedge \s(p_{1})
\wedge \neg \Diamond \s(p_{1}).
$$
Hence, by Claim~\ref{closer},
$$
x \models \s(\pi_{1}) \quad \text{iff} \quad x \models (\Diamond \alpha_{0}^{1}
\vee \alpha_{0}^{1}) \wedge \neg \Diamond \alpha_{0}^{0} \wedge \neg \Diamond
\alpha_{0}^{2} \wedge \overline{\alpha}_{k_{i}}^{1} \wedge \neg \Diamond
\overline{\alpha}_{k_{i}}^{1}.
$$
(Observe that $\s(p_{1})$ occurs within the scope of a $\Diamond$. Hence, we
obtain this equivalence only for points of distance $\le 5$ from $x_{0}$.) But
this is equivalent to $x \models \overline{\alpha}_{k_{i}}^{1}$.
\end{proof}
\begin{claim}\label{again2}
For all $x$ of distance $\le 4$ from $x_{0}$, {\rm (i)} $x \models \s(\pi_{2})$
iff $z \models \overline{\alpha}_{k_{i}+1}^{1}$, and {\rm (ii)} $x \models
\s(\tau_{2})$ iff $z \models \overline{\alpha}_{l_{i}+1}^{2}$.
\end{claim}
\begin{proof}[of Claim]
We only prove (i). Suppose $x$ is given. We know that
$$
\s(\pi_{2}) ~=~ \Diamond \alpha_{0}^{1} \wedge \neg \Diamond \alpha_{0}^{0}
\wedge \neg \Diamond \alpha_{0}^{2} \wedge \Diamond \s(p_{1}) \wedge \neg
\Diamond^{2} \s(p_{1}).
$$
Hence, by Claim~\ref{closer},
$$
x \models \s(\pi_{2}) \quad \text{iff} \quad x \models \Diamond \alpha_{0}^{1}
\wedge \neg \Diamond \alpha_{0}^{0} \wedge \neg \Diamond \alpha_{0}^{2} \wedge
\Diamond \overline{\alpha}_{k_{i}}^{1} \wedge \neg \Diamond^{2}
\overline{\alpha}_{k_{i}}^{1}.
$$
(In this case $\s(p_{1})$ occurs within the scope of a $\Diamond^{2}$.
Therefore, we obtain this equivalence for points $x$ of distance $\le 4$ from
$x_{0}$.) But this formula is in fact the definition of
$\overline{\alpha}^{1}_{k_{i}+1}$.
\end{proof}

As in the proof of Lemma~\ref{main-lemma}, we now make a case distinction
according to rule $I_{i+1}$ used to transform $\langle t_i,k_i,l_i\rangle$ to
$\langle t_{i+1},k_{i+1},l_{i+1}\rangle$. Here we only consider the case of
$I_{i+1} = t_i \to \langle t_{i+1},1,0\rangle$, and leave the remaining three
cases to the reader. We need to show that
\begin{itemize}
\item[(a)] $x_{0} \models  \s (\exists \varepsilon(t_{i},\pi_{1},\tau_{1})$) and

\item[(b)] $x_{0} \not\models  \s (\exists
\varepsilon(t_{i+1},\pi_{2},\tau_{1}))$,
\end{itemize}
which, as before, would imply $x_{0}\models \s(\psi({\mathfrak b}))$.

\smallskip

(a) As $x_{0} \models \exists \varepsilon(t_{i}, \alpha_{k_{i}}^{1},
\alpha_{l_{i}}^{2})$, we have some $z$ such that $x_{0} S n^{\mathfrak V} S z$
and
$$
z \models \Diamond \alpha_{t_{i}}^{0}\wedge\neg\Diamond
\alpha_{t_{i}+1}^{0}\wedge\Diamond \alpha_{k_{i}}^{1} \wedge\neg\Diamond^2\alpha_{k_{i}}^{1}\wedge
\Diamond \alpha_{l_{i}}^{2}\wedge\neg\Diamond^2\alpha_{l_{i}}^{2}.
$$
Clearly, it is sufficient to show
$$
z \models \Diamond \alpha_{t_{i}}^{0}\wedge\neg\Diamond \alpha_{t_{i}+1}^{0}
\wedge \Diamond \s (\pi_{1}) \wedge \neg \Diamond^{2} \s(\pi_{1}) \wedge
\Diamond \s (\tau_{1}) \wedge \neg \Diamond^{2} \s (\tau_{1}).
$$
But this follows from Claim~\ref{again1}: just observe that $z$ is of distance
$\le 2$ from $x_{0}$, while $\s(\pi_{1})$ and $\s(\tau_{1})$ occur within the
scope of $\Diamond^{2}$.

\smallskip

(b) To show $x_{0} \not\models \s(\exists \varepsilon(t_{i+1},
\pi_{2},\tau_{1}))$, suppose otherwise. Then there is $z$ such that $x_{0} S
n^{\mathfrak V} S  z$ and
$$
z \models
\varepsilon(t_{i+1}, \s(\pi_{2}), \s(\tau_{1})).
$$
This means that
$$
z \models \Diamond \alpha_{t_{i+1}}^{0}\wedge\neg\Diamond
\alpha_{t_{i+1}+1}^{0}\wedge\Diamond \s(\pi_2) \wedge\neg\Diamond^2 \s(\pi_2)
\wedge \Diamond \s(\tau_1) \wedge\neg \Diamond^2\s(\tau_1).
$$
By Claims~\ref{again1} and \ref{again2} this implies
$$
z \models \Diamond \alpha_{t_{i+1}}^{0}\wedge\neg\Diamond
\alpha_{t_{i+1}+1}^{0}\wedge\Diamond \alpha_{k_{i}+1}^{1} \wedge\neg\Diamond^2 \alpha_{k_{i}+1}^{1}
\wedge \Diamond \alpha_{l_{i}}^{2} \wedge\neg \Diamond^2 \alpha_{l_{i}}^{2}.
$$
It follows that
$$
z \models \varepsilon(t_{i+1}, \alpha_{k_{i}+1}^{1}, \alpha_{l_{i}}^{2})
$$
and we arrive at a contradiction, because  $\alpha_{k_{i}+1}^{1}=
\alpha_{k_{i+1}}^{1}$.

This completes the proofs of Lemma~\ref{lem3} and Theorem~\ref{main2}.
\end{proof}


\section{Applications to description logics}

In this section, we briefly comment on the consequences of our results in the
context of description logics \cite{DLHandbook}. We remind the reader that
description logics (DLs, for short) are knowledge representation and reasoning
formalisms in which complex concepts are defined in terms of atomic concepts
using certain constructors. DLs are then used to represent, and reason about,
various relations between such complex concepts (typically, the subsumption
relation). The basic Boolean description logic $\mathcal{ALC}$ has as its
constructors the Boolean connectives and the universal restriction $\forall r$,
which, for a concept $C$ and a binary relation symbol $r$, gives the concept
$\forall r.C$ containing precisely those objects $x$ from the underlying domain
for which $y\in C$ whenever $xry$. The language $\mathcal{ALC}$ is a notational
variant of the basic modal logic {\ml K} with infinitely many modal operators:
propositional variables correspond to atomic concepts, while $\forall r.C$ is
interpreted in a relational structure in the same way as $\Box_{r}$ (the modal
box interpreted by the accessibility relation $r$). We refer the reader to
\cite{DLHandbook} for precise definitions and a discussion of syntax and
semantics of $\mathcal{ALC}$ and other description logics.

It has been argued in \cite{Baader&Narendran} that for many applications of DLs
it would be useful to have an algorithm capable of deciding, given two complex
concepts $C_{1}$ and $C_{2}$, whether there exists a substitution $\s$ (of
possibly complex concepts in place of atomic ones) such that $\s(C_{1})$ is
equivalent to $\s(C_{2})$ in the given DL.\footnote{This is the simplest version
of the decision problem they consider. More generally, Baader and Narendran
\citeyear{Baader&Narendran} consider the problem whether there exists such a
substitution which leaves certain atomic concepts intact. We will not consider
this more complex decision problem in this paper.} We call this problem the
\emph{concept unification problem}. A typical application of such an algorithm
is as follows. In many cases, knowledge bases (ontologies) based on DLs are
developed by different knowledge engineers over a long period. It can therefore
happen that some concepts which, intuitively, should be equivalent, are
introduced several times with slightly different definitions. To detect such
redundancies, one can check whether certain concepts can be unified.
Unifiability does not necessarily mean that these concepts have indeed been
defined to denote the same class of objects---but this fact can serve as an
indicator of a possible redundancy, so that the knowledge engineer could then
`double check' the meaning of those concepts and change the knowledge base
accordingly.

The concept unification problem for $\mathcal{ALC}$ is easily seen to be
equivalent to the unification problem for the modal logic ${\ml K}$ with
infinitely many modal operators: formulated for the modal language, the problem
is to decide whether, given two modal formulas $\varphi_{1}$ and $\varphi_{2}$,
there exists a substitution $\s$ such that, for every Kripke model $\mathfrak M$
and every point $x$ in it,
$$
(\mathfrak M,x) \models \s(\varphi_{1}) \quad \mbox{ iff } \quad (\mathfrak M,
x) \models \s(\varphi_{2}).
$$
This is obviously equivalent to the validity of $\s(\varphi_{1} \leftrightarrow
\varphi_{2})$. Baader and Kuesters \citeyear{BaaderKuesters-LPAR} and Baader and
Narendran \citeyear{Baader&Narendran} develop decision procedures for the
concept unification problem for a number of sub-Boolean DLs, that is, DLs which
do not have all the Boolean connectives as constructors and are, therefore,
either properly less expressive than $\mathcal{ALC}$ or incomparable with
$\mathcal{ALC}$. The investigation of the concept unification problem for
Boolean DLs, that is, $\mathcal{ALC}$ and its extensions, is left as an open
research problem.

It should be clear that we have to leave open the decidability status for the
concept unification problem for $\mathcal{ALC}$ as well. However, we obtain the
undecidability of this problem for extensions of $\mathcal{ALC}$ with nominals.
In contemporary description logic research and applications, nominals play a
major role, see e.g., \cite{HorrocksSattler2005} and references therein. The
smallest description logic containing $\mathcal{ALC}$ and nominals is known as
$\mathcal{ALCO}$, and by extending the mapping between modal and description
languages indicated above, one can see that $\mathcal{ALCO}$ is a
straightforward notational variant of the modal logic with infinitely many modal
operators and nominals. Hence, as a consequence of Theorem~\ref{main2} we
obtain:

\begin{theorem}
The concept unification problem for $\mathcal{ALCO}$ is undecidable.
\end{theorem}

Moreover, the undecidability proof goes through as well for extensions of
$\mathcal{ALCO}$ such as, for example, $\mathcal{ALCQO}$ and $\mathcal{SHIQO}$,
the description logic underlying {\ml OWL-DL} \cite{HoPH03a}.

Another family of description logics for which the concept unification problem
turns out to be undecidable are those extensions of $\mathcal{ALC}$ where the
universal role is definable. The minimal description logic of this sort, widely
used in DL applications, is known nowadays as $\mathcal{SHI}$. Originally,
Horrocks and Sattler \citeyear{Horrocks98j} introduced this logic under the name
$\mathcal{ALCHI}_{R^+}$. In $\mathcal{SHI}$, the signature of $\mathcal{ALC}$ is
extended by

\begin{itemize}
\item infinitely many relation symbols, which are interpreted by
\emph{transitive relations},

\item and for each relation symbol $r$, there is a relation symbol
$r^{-}$, which is interpreted by the inverse of the interpretation of $r$.
\end{itemize}
The concept unification problem for $\mathcal{SHI}$ remains open. However, when
considering $\mathcal{SHI}$ it is not the concept unification problem one is
mainly interested in, but its generalisation to the \emph{concept unification
relative to role axioms}\footnote{In description logic, the most useful
generalisation of the concept unification problem is unification relative to
TBoxes \emph{and} RBoxes. We will not discuss this generalisation here because
the undecidability results presented in this paper trivially hold for it as
well.}: in $\mathcal{SHI}$ and its extensions one can state in a so-called RBox
(role box) that the interpretation of a relation symbol $r$ is included in the
interpretation of a relation symbol $s$, in symbols $r \sqsubseteq s$. Now,
$\mathcal{SHI}$ concepts $C$ and $D$ are called unifiable relative to an RBox
$R$ iff there exists a substitution $\s$ (of complex $\mathcal{SHI}$-concepts
for atomic ones) such that $\s(C)$ is equivalent to $\s(D)$ in every model
satisfying the RBox $R$. It easily seen that this problem is undecidable.
Indeed, consider the RBox $R$ consisting of $s \sqsubseteq s^{-}$, $s^{-}
\sqsubseteq s$, and $r \sqsubseteq s$, where $s$ is a transitive role. Then, in
every model for $R$, $s$ is transitive, symmetric and contains $r$. By replacing
the operator $\Box$ with $\forall r$ and the operator $\forall$ with $\forall s$
in the proof of Theorem~\ref{main1}, one can easily show that concept
unification relative to the RBox $R$ is undecidable. Thus we obtain the
following:

\begin{theorem}
The concept unification problem relative to role axioms for $\mathcal{SHI}$ is undecidable.
\end{theorem}

This undecidability proof also goes through for extensions of $\mathcal{SHI}$
such as, for example, $\mathcal{SHIN}$ and $\mathcal{SHIQ}$.


\section{Conclusion}

In this paper, we have shown that for two standard constructors of modal
logic---the universal modality and nominals---the unification and admissibility
problems are undecidable. It follows that both unification and admissibility are
undecidable for all standard hybrid logics and many of the most frequently
employed description logics.

Many intriguing problems remain open. The question whether the unification and
admissibility problems for ${\ml K}$ (or, equivalently, $\mathcal{ALC}$) are
decidable is one of the major open problems in modal and description logic.


\begin{acks}
We were partially supported by the U.K.\ EPSRC grants GR/S61966, GR/S63182,
GR/S63175, GR/S61973.
\end{acks}



\begin{received}
...
\end{received}
\end{document}